\documentclass[preprint,review,12pt]{elsarticle}

\usepackage{amssymb}
\usepackage{amsmath}

\usepackage{booktabs,tabularx}

\journal{Thermochimica Acta}

\usepackage{lineno}
\usepackage{xcolor}

\begin{document}

\begin{frontmatter}



\title{Thermal response of an in-situ STEM MEMS chip under rapid pulse heating}

\author[add1]{Phillip Dumitraschkewitz}
\ead{phillip.dumitraschkewitz@unileoben.ac.at}
\author[add1]{Thomas Kremmer}
\ead{thomas.kremmer@unileoben.ac.at}

\affiliation[add1]{organization={Department Metallurgy, Chair of Nonferrous Metallurgy, Montanuniversit\"at Leoben},
            addressline={Franz-Josef-Str. 18}, 
            city={Leoben},
            postcode={8700}, 
            state={},
            country={Austria}}

\begin{abstract}
In-situ rapid solidification studies demand measurements of thermal histories with high temporal resolution. We present a simple, effective setup to quantify the cooling response of an uncoated commercial Protochips Fusion MEMS chip in an in-situ scanning transmission electron microscopy (STEM) context. We drive user-defined temperature programs via an arbitrary waveform generator (AWG), while recording the voltage drops across a series shunt to reconstruct chip resistance and temperature at sub-millisecond resolution. We confirm the response times inferred from the current; however, the temperature obtained from the physically linked resistance, $T(R)$, evolves more slowly. Analysis of the maximum cooling step reveals an exponential-like relaxation with time constant $\tau=1.80$ ms, consistent with reported thermal lag constants for fast scanning calorimetry. From the time to reach $95\%$ of the temperature difference $\Delta T$, we measure an average cooling rate of $\approx 7.9\times 10^{4}$ K/s. Robustness checks include repeated $R(T)$ measurements (revealing a modest downward drift approaching an asymptote), a 10 k$\Omega$ test load, and characterization of small off-duty arbitrary waveform generator leakage/offsets. These findings define practical bounds on achievable thermal-path rates when planning in-situ electron microscopy experiments with this chip platform.
\end{abstract}

\begin{graphicalabstract}
\includegraphics{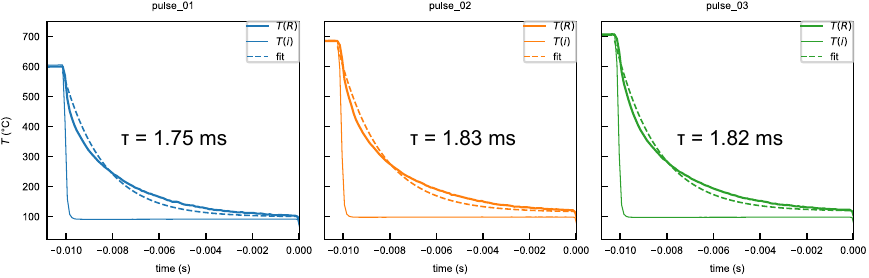}
\end{graphicalabstract}

\begin{highlights}
\item Customized measurement setup: arbitrary waveform generator and a digital oscilloscope
\item Current drop times are verified to be $<$1 ms.
\item Maximum cooling step exhibits exponential relaxation with $\tau = 1.80\,\mathrm{ms}$
\item Average cooling rate is $\approx 7.9\times10^{4}\ \mathrm{K/s}$.
\end{highlights}


\end{frontmatter}



\section{Introduction}

With the advancement of additive manufacturing methods~\cite{debroyAdditiveManufacturingMetallic2018}, efforts to study rapid solidification -- and its direct correlation between thermal history and microstructure at small scales -- have likewise increased.
Several studies focus on fast scanning calorimetry (FSC)~\cite{schickFastScanningCalorimetry2016,gaoNanocalorimetryDoorOpened2019} with powder particles and additional ex-situ characterization: investigating the undercooling–microstructure correlation~\cite{yangNucleationBehaviourMicrostructure2021, pengContinuousCoolingIsothermal}, studying the influence of powder modifications on undercooling and microstructure~\cite{zhuravlevAssessmentAlZnMgCuAlloy2021}, or quantifying changes in specific heat capacity~\cite{quickFastDifferentialScanning2023} of rapidly solidified particles.

In another approach, in situ transmission electron microscopy (TEM) experiments have been performed.~\cite{mckeownSituTransmissionElectron2014, mckeownInsituTransmissionElectron2013, mckeownTimeResolvedSituMeasurements2016} In these, sections of a thin film were melted using a pulsed laser, and the velocity of the liquid/solid interface during solidification was measured with dynamic TEM -- a technique that enables imaging on the microsecond timescale during solidification. In addition, multiple post-solidification STEM/TEM techniques have been applied to investigate composition and crystallography. However, direct experimental temperature measurements are currently not available.

In our earlier work on rapid solidification~\cite{dumitraschkewitzMEMSBasedSituElectronmicroscopy2022}, we used a chip-based setup in an in situ STEM experiment with both pulse heating and linear (Joule) heating.

\begin{figure}
    \centering
    \includegraphics[width=\linewidth]{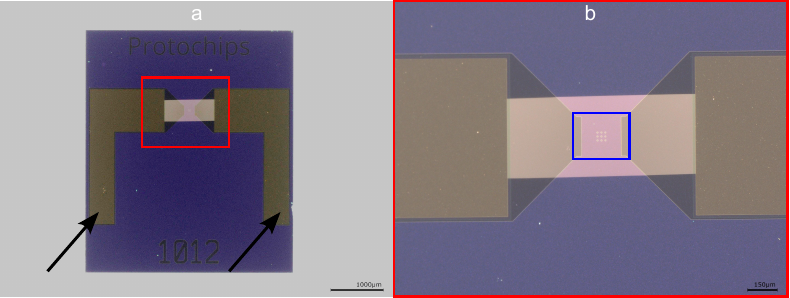}
    \caption{Overview of the used in situ STEM chip system. a Light optical microscope image of a whole chip, black arrows mark electrical contact area for the pins of the holder, red rectangle shows the area of the magnification of b. b Pink area (low conductivity ceramic~\cite{allardNewMEMSbasedSystem2009}) within the blue rectangle is the Joule heated area of the chip.}
    \label{fig:overview_chip}
\end{figure}

When we compare in situ STEM/TEM chips to typical FSC chips~\cite{schickFastScanningCalorimetry2016}, we see several differences. While FSC chips use thermocouples for temperature measurements, the here investigated system of in situ TEM chips lack an independent temperature sensor decoupled from the power source; instead, resistance or current is measured~\cite{allardNewMEMSbasedSystem2009}. As a result, FSC chips generally offer higher temperature accuracy. The in situ STEM chip is engineered to minimize mechanical drift under thermal loading. Membranes of FSC chips are usually thicker; for example, the commercial UFS chip~\cite{manualThermalAnalysisExcellence} has a much thicker membrane ($\sim 2\,\mu\mathrm{m}$) than the electron-transparent membrane of the in situ STEM chip ($\approx 120\,\mathrm{nm}$~\cite{protochipsinc.FusionSelectUser2020}). An in situ chip system of the type used in this study is shown in Fig.~\ref{fig:overview_chip}. The chip's membrane material is SiC-based~\cite{US9048065}, and the Joule-heated area is approximately $200\times 200\,\mu\mathrm{m}$; see the blue rectangle in Fig.~\ref{fig:overview_chip}b. For a more detailed description of the in situ STEM chip system design, the reader is referred to Ref.~\cite{allardNewMEMSbasedSystem2009}. Typical sample masses are higher for FSC experiments, with thicknesses in the micron range, whereas STEM samples must usually be smaller than one micron -- typically below $500\,\mathrm{nm}$ for electron transparency~\cite{carterTransmissionElectronMicroscopy2016}. In addition, the heat-flow conditions differ: FSC is typically performed in a cooled, purged inert-gas atmosphere, whereas STEM experiments are carried out under high-vacuum conditions ($\approx 8 \times 10^{-6}$ Pa~\cite{dumitraschkewitzMEMSBasedSituElectronmicroscopy2022}).

We estimate the radiative heat transfer coefficient using Eq.~\eqref{eq:radiation}~\cite{minakovThermalContactConductance2020}:
\begin{align}
    \alpha_r = \epsilon \sigma (T^4-T_\mathrm{RT})/(T-T_\mathrm{RT}), \label{eq:radiation}
\end{align}
with $\epsilon = 0.2$, $\sigma = 5.67\times10^{-8}\ \mathrm{W\,m^{-2}\,K^{-4}}$, $T$ set to the maximum sample temperature from Table~\ref{tab:parameters}, and $T_\mathrm{RT}$ as room temperature, and compare the results to typical solid–solid contact conductance values. Using these parameters gives $\alpha_r \approx 13\ \mathrm{W\,m^{-2}\,K^{-1}}$, which is orders of magnitude below a representative solid–solid contact conductance of $2.3\times10^{3}\ \mathrm{W\,m^{-2}\,K^{-1}}$~\cite{minakovThermalContactConductance2020}. Thus, for in situ STEM experiments in vacuum, chip cooling is dominated by in‑plane conduction of the membrane. In contrast, during FSC experiments, the gas temperature, flow, and composition can be adjusted to modify the cooling conditions and thereby the maximum achievable cooling rates.

Regarding the possible heating or cooling rates for the chips used in our study~\cite{dumitraschkewitzMEMSBasedSituElectronmicroscopy2022}, Ref.~\cite{allardNewMEMSbasedSystem2009} states that temperature cycling with $>10^6\,\mathrm{K/s}$ can be achieved, while a later patent~\cite{US9048065} claims cooling rates $>10^4\,\mathrm{K/s}$.

In open-loop mode, heating/cooling rates are software limited to a maximum of $10^6\,\mathrm{K/s}$. However, the standard integration time of the commercial measurement setup is $100\,\mathrm{ms}$~\cite{dumitraschkewitzMEMSBasedSituElectronmicroscopy2022}; therefore, the temperature profile of a rapid heating pulse cannot be tracked. In closed-loop mode, the software limits the rate to as low as $10\,\mathrm{K/s}$~\cite{protochipsinc.FusionSelectUser2020}.

To examine the relationship between the thermal path and microstructure of materials with this method, it is essential, in the first place, to be able to measure the thermal path with sufficiently high temporal resolution and, further, to have a detailed understanding of the chip’s thermal behavior and limitations.

Therefore, this paper investigates the cooling behavior of a commercial Protochips Fusion chip (Morrisville, USA) using the characteristic resistivity over temperature $R(T)$ relationship~\cite{allardNewMEMSbasedSystem2009} of its SiC-based~\cite{US9048065} heated membrane.
A simple yet effective measurement setup is used to investigate the response to a rapid cooling step applied to the chip and to discuss the robustness of the measurements.

\section{Method \label{sec:method}}

In this section the customized measurement setup is described in detail: derivation of the used temperature-resistance relation, the physical realization of the setup, and control and creation of the time-temperature profiles.

As a first step, empty‑chip experiments are performed using the original setup and calibration. Fast‑rate experiments are then conducted with a customized setup whose main components are an arbitrary waveform generator (AWG; Keithley 6221, Cleveland, USA) and a digital oscilloscope (Tektronix MSO 44~\cite{tektronix4SeriesMSO2021}, Beaverton, USA).

\subsection{Temperature–resistance relationship\label{sec:calibration_curve}}

The following procedure was applied to calculate the current ($i$)-temperature ($T$) and resistance ($R$)-temperature relationships, which are subsequently used in the customized setup:

The chip, without a sample, is inserted into the Protochips AXON Fusion holder, and the holder is inserted into the TEM. In closed‑loop mode, the commercial setup controls the current via a feedback loop that measures the chip’s resistance and maps it to temperature using a calibration, whereas in open‑loop mode only the current is controlled and used to estimate the temperature~\cite{allardNewMEMSbasedSystem2009, protochipsinc.FusionSelectUser2020}.

With the commercial setup an open-loop staircase temperature program is run with $10\,\mathrm{ms}$ heating between isothermal segments of $300\,\mathrm{ms}$. A total of 145 steps is used in the range from $25$ to $750\,^\circ\mathrm{C}$. The measurement is a two-point measurement. The experimental logfile (recorded with Protochips Clarity software), which in addition to time, temperature, and current also stores the measured resistance values for the staircase temperature program, is used to compute the $i\!\to\!T$ and $R\!\to\!T$ relationships. For each isothermal segment, typically three measurement points are acquired (at $100\,\mathrm{ms}$ intervals). For the heating and cooling segments, each measurement value for a given temperature is averaged. For some chips at low temperatures ($\lesssim 60\,^\circ\mathrm{C}$), the $R\!\to\!T$ relationship is not bijective; the values in this region are replaced by averages to obtain a bijective curve, but these temperatures are neglected in the analysis. For the chip used in the experiments reported here, $R(T)$ is monotonically decreasing.

\begin{figure}
    \centering
    \includegraphics[width=\linewidth]{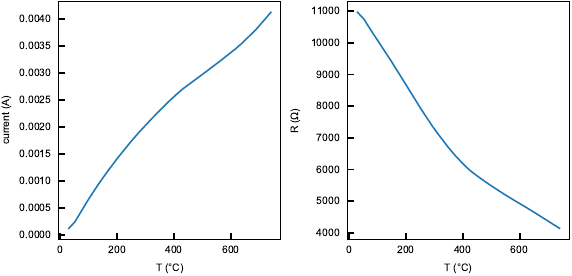}
    \caption{Temperature $(T)$--current $(i)$ and temperature $(T)$--resistance $(R)$ relationships.}
    \label{fig:calibration_relationship}
\end{figure}

Generally, the measured resistance values show a small hysteresis upon heating and cooling.
Therefore, for each temperature data point, the resistance values from the heating and cooling segments are interpolated and then averaged. The scatter in the measurements, quantified by the standard deviation, is typically $<50\,\Omega$ at low temperature and $<5\,\Omega$ at high temperature; the scatter decreases with increasing temperature. The $T$--$i$ and $T$--$R$ curves used are shown in Fig.~\ref{fig:calibration_relationship}.
The reported logfile values of the isothermal segments follow the Ohmic power law in Eq.~\ref{eq:power_resistivity}.

\begin{align}
    p &= R i^2 \label{eq:power_resistivity}
\end{align}

\subsection{Measurement setup}

\begin{figure}
    \centering
    \includegraphics[width=\linewidth]{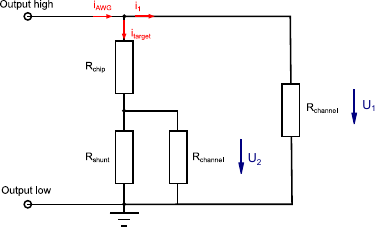}
    \caption{Schematic circuit diagram for the measurement setup, simplified to Ohmic resistances.}
    \label{fig:circuit}
\end{figure}

The setup presented here is similar to the measurement principle of Ref.~\cite{dirisaglikHighSpeedHigh2015}. A simplified circuit diagram of the used measurement setup is shown in Fig.~\ref{fig:circuit}. A power resistor (shunt resistance, $R_{\text{shunt}}$) is connected in series with the chip. The oscilloscope measures the total voltage across chip plus shunt on channel 1 ($U_1$) and the voltage across the shunt on channel 2 ($U_2$). Both channels share a common ground, which is connected to the AWG instrument ground and to Output Low on the triax connector~\cite{instrumentsincModel6220DC2008}. Component specifications are listed in Tab.~\ref{tab:specifications}.

\begin{table}[t]
  \centering
  \small
  \setlength{\tabcolsep}{6pt}
  \begin{tabularx}{\linewidth}{@{} l l >{\raggedleft\arraybackslash}X >{\raggedleft\arraybackslash}X @{}} 
    \toprule
    Part & Type/Comment & Value/Range & Accuracy \\
    \hline
    AWG & Keithley 6221~\cite{instrumentsincModel6220DC2008} & & \\
        & Amplitude offset & $<20$ mA & $0.2\%$ rdg. + $40\,\mu$A \\
    \hline
    Digital oscilloscope & Tektronix MSO 44~\cite{tektronixinc.4SeriesMSO2020} & & \\
        & Channel resistance, $R_\mathrm{channel}$ & $250\ \mathrm{k}\Omega$ & \\
        & DC gain accuracy & & $1\%$ \\
        & Offset accuracy & for 1 V/div & 200 mV \\
    \hline
    Shunt resistance, $R_\mathrm{shunt}$ & Power resistor 5 W & $2.2\ \mathrm{k}\Omega$ & $5\%$ \\
    \hline
    Test resistance & Power resistor & $10\ \mathrm{k}\Omega$ & $10\%$ \\
    \bottomrule
  \end{tabularx}
  \caption{Summary of part specifications.}
  \label{tab:specifications}
\end{table}

The raw measurement values are averaged over time by numerically integrating them over a time interval (“integration time”) using the trapezoidal rule (assuming a linear relationship between measurement points) and dividing the integral by the interval length. The chip resistance is then calculated from these values using Eq.~\ref{eq:resistivity_setup}.

\begin{align}
    R_{\text{shunt}}\parallel R_\text{channel} &= \frac{R_\text{channel}\,R_{\text{shunt}}}{R_\text{shunt}+R_\text{channel}}\label{eq:parallel_R}\\
    R_{\text{chip}} &= \frac{R_{\text{shunt}}\parallel R_\text{channel}}{U_2}\,\bigl(U_1-U_2\bigr) \label{eq:resistivity_setup}
\end{align}

\subsection{Creation of the temperature program}

\begin{figure}
    \centering
    \includegraphics[width=\linewidth]{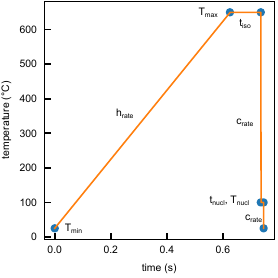}
    \caption{Schematic of the parameterized temperature program for the customized setup. Blue dots represent the time temperature pairs that define the temperature program.}
    \label{fig:temp_programs_schematic}
\end{figure}

To run a temperature program, an $i_\text{target}(t)$ waveform is precomputed from the $T\leftrightarrow i$ relationship (Fig.~\ref{fig:calibration_relationship}). The time--temperature program of Fig.~\ref{fig:temp_programs_schematic} is transformed into $i_\text{target}(t)$.

Due to the additional resistance in parallel with the chip resistance (Fig.~\ref{fig:circuit}), the applied current from the AWG ($i_\text{AWG}$) must be increased as in Eq.~\ref{eq:i_increase}.

\begin{align}
    i_1 &= \frac{\bigl(R(T)+R_\text{shunt}\parallel R_\text{channel}\bigr)\,i_\text{target}}{R_\text{channel}}\label{eq:i_increase}\\
    i_\text{AWG} &= i_\text{target}+i_1 \label{eq:i_AWG}
\end{align}

The $i_\text{target}(t)$ program consists of 65535 points on equally spaced time steps, which are loaded onto the Keithley 6221. A phase marker is set at the end of the cycle, and a single cycle is run. The phase marker triggers the oscilloscope. The phase-marker output is connected to the AUX input of the oscilloscope.

The customized control uses a parameterized temperature program with the parameters $T_{\min}$, $T_{\max}$, $T_{\text{nucl}}$, $t_{\text{iso}}$, $t_{\text{nucl}}$, $h_{\text{rate}}$, and $c_{\text{rate}}$, which are either directly defined or computed from $(T_{\max}-T_{\text{nucl}})/\Delta t$, where $\Delta t$ is limited by the smallest realizable time step of the AWG. The temperature program was initially designed to cool from the melt to a defined nucleation temperature, $T_{\text{nucl}}$. In Fig.~\ref{fig:temp_programs_schematic}, the parameterized temperature program is shown.

\section{Experimental conditions}

Following Ref.~\cite{tunesFastImplantationfreeSample2021}, an electron‑transparent specimen was cut with a scalpel, transferred using a hair, and placed onto the chip membrane without further additional mounting. The alloy has a nominal composition of 3.0~wt.\% Fe, 0.14~wt.\% Si, with the balance Al. The used chip is an uncoated chip with environmental calibration. Here, ``environmentally calibrated'' means the chip was also calibrated by the manufacturer for use in an atmospheric environment, such as an environmental TEM. The chip was mounted on the Protochips Fusion AX holder, and loaded into a Thermo Fisher Scientific\textsuperscript{TM} Talos F200X G2 scanning transmission electron microscope. Experiments were conducted under high-vacuum ($\approx 8 \times 10^{-6}$ Pa~\cite{dumitraschkewitzMEMSBasedSituElectronmicroscopy2022}), in STEM mode with a field-of-view of 23.7 $\mu$m, with an image size of 256$\times$256 pixel, a spot size setting of ``6'', a dwell time of 750 nm, C1 aperture of 2000 $\mu$m and C2 of 50 $\mu$m.

The parameterized temperature programs were set with the parameters shown in Table~\ref{tab:parameters}, transformed to $i_{\text{target}}$, and applied via the AWG.

\begin{table}
    \centering
    \begin{tabular}{cccccccc}
         pulse \#& $T_{\min}$ & $T_{\max}$ & $T_{\text{nucl}}$ & $h_{\text{rate}}$ & $c_{\text{rate}}$ & $t_{\text{iso}}$ & $t_{\text{nucl}}$\\
         \hline
         01 &25 & 650 & 100 & 1000 & 6250000 & 110 & 10\\
         02 &25 & 720 & 100 & 1000 & 3475000 & 110 & 10\\
         03 &25 & 750 & 100 & 1000 & 3625000 & 110 & 10\\
    \end{tabular}
    \caption{Experimental parameters for the parameterized temperature program. Temperatures ($T$) are given in °C and times ($t$) in ms, rates in K/s.}
    \label{tab:parameters}
\end{table}

\section{Results and discussion}

\subsection{Maximum cooling\label{sec:max_cooling}}

\begin{figure*}
    \centering
    \includegraphics[width=\linewidth]{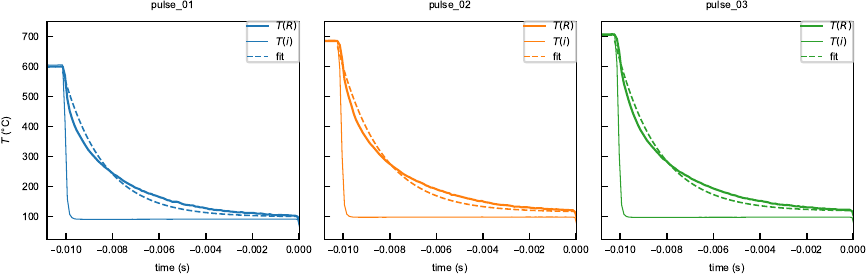}
    \caption{Thermal response of the cooling segment of the temperature programs. $T(i)$ is the temperature obtained from the $i$–$T$ relationship, $T(R)$ from the $R$–$T$ relationship, and the orange curve is an exponential fit to the maximum-cooling segment of $T(R)$. An integration time of $0.1\,\mathrm{ms}$ is used for the shown data.}
    \label{fig:max_cooling}
\end{figure*}

In Fig.~\ref{fig:max_cooling}, the cooling segment (see also Fig.~\ref{fig:temp_programs_schematic}) is examined in detail. The temperature is calculated from the $T(i)$ or $T(R)$ curves. The current -- and correspondingly $T(i)$ -- drops within less than $1\,\mathrm{ms}$. 

However, $T(R)$ captures the material’s temperature dependence via its electrical resistance, which is directly related to its specific resistivity through the chip’s dimensions. The $T(i)$ curve corresponds to temperature only for sufficiently long time spans, when the input Joule heating power balances the heat losses to the environment and a stationary temperature field is established.

A fit to an exponential decay function (Eq.~\ref{eq:exp_decay}), where $T_0$ is the high-step temperature, $T_\text{nucl}$ the low-step value, and $t_0$ the segment start time. $T_0$ is measured by the mean value from of the last 10 ms of the high-step value $T(R)$, $T_\text{nucl}$ is estimated by the last value of $T(R)$ within the cooling segment. The resulting values, together with the fit results, are listed in Tab.~\ref{tab:results_org}. This yields an exponential time constant (Eq.~\ref{eq:thermal_lag}) $\tau$ of $1.80\pm0.04\,\mathrm{ms}$ (sample standard deviation of the three fitted parameter values). This $\tau$ can be compared to the thermal-lag constant of FSC data, which gives similar values: $2.2\pm0.6\,\mathrm{ms}$ in Ref.~\cite{schaweTemperatureCorrectionHigh2021} and $2.3\,\mathrm{ms}$ in Ref.~\cite{pogatscherInsituProbingMetallic2014}.

\begin{align}
    \Delta T &= T_0 - T_\text{nucl}\\
    T(t) &= T_0 - \Delta T\left(1 - e^{-a (t-t_0)}\right) \label{eq:exp_decay}\\
    \tau &= 1/a \label{eq:thermal_lag}
\end{align}

\begin{table}
    \centering
    \begin{tabular}{crrrrrr}
        pulse \#&$T_0$ &  $T_\mathrm{nucl}$&  $\Delta T$&  $\tau$&  $\sigma_\mathrm{fit}$& $c_\mathrm{rate, 95\%}$\\
        \hline\hline
        01& 599& 99& 500& 1.75& 0.03& 72620 \\
        02& 685& 115& 570& 1.83& 0.03& 80211 \\
        03& 705& 118& 587& 1.82& 0.03& 84248 \\
    \end{tabular}
    \caption{Measured data during the fit procedure for uncorrected data. Temperatures are given in °C, temperature difference $\Delta T$ in K, the average cooling rate $c_\mathrm{rate, 95\%}$ in K/s, the relaxation time and $\sigma_\mathrm{fit}$ in ms. The mean value of the relaxation time and the sample standard deviation of the measurement values is $\tau=1.80\pm0.04$ ms.}
    \label{tab:results_org}
\end{table}

\subsection{Heating influence of the electron beam\label{sec:beam_heating}}

Following Refs.~\cite{egertonRadiationDamageTEM2004,egertonControlRadiationDamage2013} (Eq.~\ref{eq:egerton}) we estimate the temperature rise $\Delta T_\mathrm{beam}$ for a beam diameter from 1.65 nm~\cite{kleebergSituHighTemperature2024} to 23.7 $\mu$m (field-of-view), for an estimated beam current ($\approx 1.15 \times$ screen current) $I$ of 138 pA, $R_0$ of 200 $\mu$m, an inelastic mean free path of the electrons $\lambda_\mathrm{inelastic}$ of 150 nm, thermal conductivity $\kappa$ of 50 Wm$^{-1}$K$^{-1}$\cite{watariEffectGrainBoundaries2003} and with an average energy loss per electron $\left<E\right>$ of 23 eV, to range within 0.9--0.2 mK. We therefore do not expect any significant influence from beam heating under our conditions.

\begin{align}
    \Delta T_\mathrm{beam} = I \left<E\right>/\lambda_\mathrm{inelastic} (0.58+2\ln(2 R_0/d)) / (4\pi \kappa) \label{eq:egerton}
\end{align}

\subsection{Robustness of the measurement}

Two features are also evident in Fig.~\ref{fig:max_cooling}: there is almost always a small gap between $T(i)$ and $T(R)$, and the fit function does not perfectly capture the curve shape. The latter is likely due, in part, to the application of a one-dimensional model to a multi-dimensional heat-transfer problem.

\begin{figure}
    \centering
    \includegraphics[width=\linewidth]{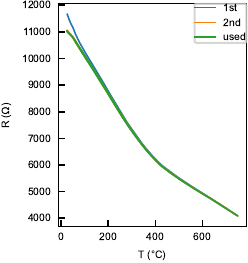}
    \caption{$R(T)$ results from initial staircase-step experiments.}
    \label{fig:initial_calibration_runs}
\end{figure}

When recording the $T(R)$ curve as outlined in Section~\ref{sec:calibration_curve}, we perform multiple repetitions, since we have observed that the measured $R(T)$ curves decrease with an increasing number of cycles (Fig.~\ref{fig:initial_calibration_runs}) but appear to approach an asymptotic limit. We attribute this behavior to prolonged storage in air and/or to annealing-induced changes in the chip membrane itself, though the exact cause is unknown to us.

No contact resistance was accounted for as we think the contribution is negligible (in the order of 10 $\Omega$~\cite{INTRODUCTIONRChipRHolder}) in comparison to the high chip resistivity itself, and also no change in mechanical load on mechanical contacts in between $T(R)$ curve measurements is expected.

\begin{figure}
    \centering
    \includegraphics[width=\linewidth]{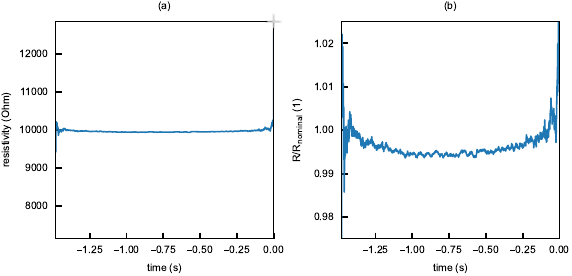}
    \caption{a) Measured resistance of a $10\,\mathrm{k}\Omega$ resistor. b) shown as the ratio of the measured to the nominal value. A current program corresponding to a chip program of symmetrical heating and cooling at $1000\,\mathrm{K/s}$ was run with $T_{\min}=25\,^\circ\mathrm{C}$, $T_{\max}=750\,^\circ\mathrm{C}$, $t_{\text{iso}}=10\,\mathrm{ms}$, $t_{\text{nucl}}=20\,\mathrm{ms}$, and $T_{\text{nucl}}=100\,^\circ\mathrm{C}$.}
    \label{fig:test_resistivity}
\end{figure}

The chip is substituted with a $10\,\mathrm{k}\Omega \pm 10\%$ power resistor, and the applied current is tested. The results are shown in Fig.~\ref{fig:test_resistivity}. The reported ratio of measured/nominal resistance is $0.9965$ on average and shows a standard deviation of $0.0198$.

\begin{figure}
    \centering
    \includegraphics[width=\linewidth]{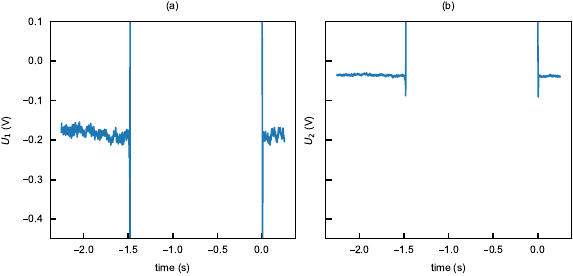}
    \caption{a) Voltages $U_1$ and b) $U_2$ showing offsets outside the AWG duty times.}
    \label{fig:offset}
\end{figure}

A potential difference is observed during the AWG off-duty time (Fig.~\ref{fig:offset}) on channels 1 and 2, which may indicate a leakage current of $\approx 16\,\mu\mathrm{A}$ (Fig.~\ref{fig:offset}) or a zero point offset of the digital oscilloscope. Before the beginning of the duty cycle, a short, even larger, negative step (corresponding to $40\,\mu\mathrm{A}$) can be seen. This offset matches well with the reported~\cite{instrumentsincModel6220DC2008} amplitude-accuracy offset of $0.2\%$ of reading $+\,0.2\%$ of the applied current range ($20\,\mathrm{mA}$), which gives $40\,\mu\mathrm{A}$.

While the temperature deduced from current and from resistance agrees well -- except for the small gap mentioned during the isothermal segments -- there is a systematic shortfall in the current through the chip relative to the setpoint, effectively leading to a $T_{\max}$ that is $\approx 50\,\mathrm{K}$ and lower. To address this, the current losses through the oscilloscope channels were compensated in later experiments in the precomputed current program, as in Eq.~\ref{eq:i_AWG}. Still, we observe a systematic shortfall in the applied current. When we virtually add the negative step value of $U_2$ seen before the AWG duty time (Fig.~\ref{fig:offset}) and recalculate the current, $i_{\text{target}}$ comes within $1.2\%$ of the intended value, which is within the amplitude magnitude accuracy in this range of 2.1\%, of the AWG~\cite{instrumentsincModel6220DC2008}.

We suppose that the AWG treats this negative step as its zero reference when sourcing current. This behavior eventually originates due to two differing offsets used by the AWG and digital oscilloscope and/or from the introduction of a ground reference into the circuit, which was required because the digital microscope uses ground as its reference. It yields a $\approx$ 3\% too low applied maximum current in the presented case, but this
does not affect the prior analysis of the pulse-run measured data.

\subsection{Estimated influence of the temperature accuracy and drift on the measurement of the relaxation time}

\begin{figure}
    \centering
    \includegraphics[width=\linewidth]{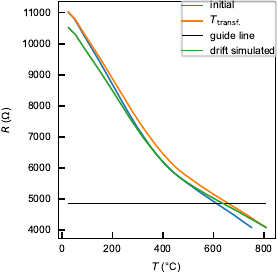}
    \caption{Changes of the $R(T)$ relation ship under transformation and simulated drift.}
    \label{fig:estimate_tau}
\end{figure}

To approximate a correction for an inaccurate $R(T)$ relationship, we apply a linear transformation to the temperature scale and assess its effect on the measured relaxation time $\tau$. Specifically, we assume that the melting temperature of pure Al at $660^\circ\mathrm{C}$ ($T_{\text{liq,Al}}$) is underestimated by $5\%$ in absolute temperature, i.e., measured at $0.95\,T_{\text{liq,Al}}$. This corresponds to a difference of $T_{\text{diff}} \approx 47\,\mathrm{K}$ between the true and observed melting points. In Fig.~\ref{fig:estimate_tau}, the temperature scale is linearly transformed according to Eq.~\ref{eq:transform_T}, changing the initial $R(T)$ relation (blue) to the transformed relation ($T_{\text{transf}}$, orange). This shifts the melting point along the drawn guide line to its correct position.

Repeating the fitting procedure with this transformed $R(T)$ relation yields no change in the resulting relaxation time $\tau$. The reason is that Eq.~\ref{eq:exp_decay} involves only temperature differences appearing in a ratio; hence, it is invariant under the applied linear transformation. This invariance is shown explicitly in Eqs.~\ref{eq:proportional_constant} and~\ref{eq:proof}, which recover Eq.~\ref{eq:exp_decay}.

\begin{align}
    T_\text{transf.} &= \frac{T_\text{liq.,Al}-T_\text{RT}}{T_\text{liq.,Al}-T_\text{diff}-T_\text{RT}}(T-T_\text{RT})+T_\text{RT} \label{eq:transform_T}\\
    b &= \frac{T_\text{liq.,Al}-T_\text{RT}}{T_\text{liq.,Al}-T_\text{diff}-T_\text{RT}} \label{eq:proportional_constant}\\
    1-e^{-(t-t0)/\tau} &= \frac{b(T(t)-T_\text{RT})+T_\text{RT}-(b(T_0-T_\text{RT})+T_\text{RT})}{(b(T_0-T_\text{RT})+T_\text{RT}-(b(T_\text{nucl}-T_\text{RT})+T_\text{RT})} \nonumber\\
    & = \frac{T(t)-T_0}{T_0-T_\text{nucl}} \label{eq:proof}
\end{align}

From Fig.~\ref{fig:initial_calibration_runs} we observe that the drop in $R(T)$ is larger at low $T$ than at high $T$, where only a small effect is visible. To test sensitivity to additional drift, we subtract a linearly decreasing resistance offset with maximum magnitude $\Delta R = 500\,\Omega$ across the full temperature range, as defined in Eqs.~\ref{eq:drift_1}--\ref{eq:drift_2}. This yields the drift-simulated relationship $R_{\mathrm{drift,sim}}(T)$, shown in green in Fig.~\ref{fig:estimate_tau}. We expect this procedure to overestimate the actual drift, particularly at high $T$, since the change from the ``2nd'' to the used $R(T)$ curve in Fig.~\ref{fig:initial_calibration_runs} is already small.

\begin{table}
    \centering
    \begin{tabular}{crrrrrr}
        pulse \#&$T_0$ &  $T_\mathrm{nucl}$&  $\Delta T$&  $\tau$&  $\sigma_\mathrm{fit}$& $c_\mathrm{rate, 95\%}$\\
        \hline\hline
        01& 621& 68& 553& 1.80& 0.04& 80344 \\
        02& 728& 85& 643& 1.83& 0.03& 89688 \\
        03& 752& 89& 664& 1.81& 0.03& 94524 \\
    \end{tabular}
    \caption{Measured data during the fit procedure for temperature corrected drift simulated data. Temperatures are given in °C, temperature difference $\Delta T$ in K, the average cooling rate $c_\mathrm{rate, 95\%}$ in K/s, the relaxation time and $\sigma_\mathrm{fit}$ in ms. The mean value of the relaxation time and the sample standard deviation of the measurement values is $\tau=1.81\pm0.01$ ms.}
    \label{tab:results_modified}
\end{table}

Repeating the fitting procedure yields $\tau = 1.81 \pm 0.01\,\mathrm{ms}$ (Tab.~\ref{tab:results_modified}), which slightly shifts the mean value and reduces the scatter among the three measurements. This suggests that drift in the $R(T)$ curve may occur between pulses. The interpretation is consistent with the slightly increasing low‑step temperature from pulse\_01 to pulse\_03 in Fig.~\ref{fig:max_cooling}: if the true $R(T)$ decreases between pulses for a given current $i(T)$, then using the uncorrected $R(T)$ for the back‑transformation overestimates $T(R)$, thereby widening the gap between $T(i)$ and $T(R)$.

\begin{align}
    T_\mathrm{normed} &= \frac{T_\text{transf.}-T_\mathrm{RT}}{T_\text{transf., max}-T_\mathrm{RT}} \label{eq:drift_1} \\
    R_\mathrm{drift, sim.} &= R(T) - \Delta R(1-T_\mathrm{normed}(T)) \label{eq:drift_2}
\end{align}

\section{Conclusions and outlook}

In summary, we verify the fast drop of the current within $<1\,\mathrm{ms}$, but the physically temperature-dependent quantity -- the resistance -- behaves differently. For a step program, an exponential fit gives a relaxation time of $1.80\pm0.04\,\mathrm{ms}$, which is very similar to values reported for FSC chips. Measuring the time for the temperature to drop to $95\%$ of $\Delta T$ yields an average cooling rate of $79000\,\mathrm{K/s}$ in the region between of approximately 700--600 to $100\,^\circ\mathrm{C}$.

A direction for future work is to use closed-loop mode initial $T(R)$ relationship measurements and validate the critical cooling rate with a metallic glass-forming alloy within the corresponding cooling rate window. Furthermore, a coupled electrothermal model (e.g., a finite‑element model) of the chip system would enhance understanding of the heat‑flow behavior.

\section*{Acknowledgments}
This work was supported with funding by Montanuniversit\"at Leoben. The TEM facility used in this work received funding from the Austrian Research Promotion Agency (FFG), project “3DnanoAnalytics”, under contract number FFG no.~858040.

The work was funded/co-funded by the European Union (ERC, HETEROCIRCAL, 101124514). Views and opinions expressed are however
those of the author(s) only and do not necessarily reflect those of the European Union or the European Research Council. Neither the European Union nor the granting authority can be held responsible for them.

\section*{Author contributions}
P.D.: Conceptualization, Funding acquisition, Formal analysis, Methodology, Investigation, Writing - original draft.\\
T.K.: Investigation, Funding acquisition, Writing - review and editing.

\section*{Data statement}
The data used in this publication are available on Zenodo in Ref.~\cite{dumitraschkewitz_2026_19202769}.

\section*{Conflict of Interest}

The authors have declared no conflict of interest.

\clearpage
\bibliographystyle{elsarticle-num}
\bibliography{paper_AlFeSi}

\end{document}